\documentclass[aps,prl,numerical,superscriptaddress,showpacs,floatfix,reprint]{revtex4-1}

\usepackage{graphicx,color}
\usepackage{amsmath,amssymb,amsfonts}
\usepackage[colorlinks=true,plainpages=true]{hyperref}

\def \be {\begin{equation}}
\def \ee {\end{equation}}
\def \ee  {\end{equation}}
\def \bea {\begin{eqnarray}}
\def \eea {\end{eqnarray}}

\def \roots{\sqrt{s_{_{NN}}}}

\def \GeVc {\mbox{$\mathrm{GeV} / c$}}

\def \gt {\mbox{$>$}}

\begin{document}
\title{
A method to test the coupling strength of the linear and nonlinear contributions to higher-order flow harmonics via 
Event Shape Engineering
}

\medskip

\author{Niseem~Magdy} 
\email{niseemm@gmail.com}
\affiliation{Department of Physics, University of Illinois at Chicago, Chicago, Illinois 60607, USA}

\author{Olga~Evdokimov} 
\affiliation{Department of Physics, University of Illinois at Chicago, Chicago, Illinois 60607, USA}

\author{Roy~A.~Lacey} 
\affiliation{Department of Chemistry, State University of New York, Stony Brook, New York 11794, USA}


\begin{abstract}
A Multi-Phase Transport (AMPT) model is used to study the efficacy of shape-engineered events to delineate the degree of coupling between the linear and nonlinear contributions to the higher-order flow harmonics $v_{4}$ and $v_{5}$. The study shows that the nonlinear contributions are strongly shape-dependent while the linear contributions are shape-independent, indicating little if any, coupling between the linear and nonlinear flow coefficients. The experimental verification of such patterns could be an invaluable tool for robust extraction of the linear and mode-coupled flow coefficients, especially for beam energies where the charged particle multiplicity and the event statistics precludes the use of current methods to establish the coupling strength.
\end{abstract}
\maketitle

\section{Introduction}

Ongoing studies of the ultra-relativistic heavy-ion collisions at the Relativistic Heavy Ion Collider and the Large Hadron Collider indicate that an exotic state of matter called Quark-Gluon Plasma (QGP) is formed in these interactions. A wealth of investigations is directed toward characterizing the dynamical evolution and the transport properties of the QGP.

Measurements of azimuthal anisotropy of particle production in heavy-ion collisions have been used in several studies to reveal the viscous hydrodynamic response of the system to the initial spatial distribution in energy density produced in the early stages of the collision~\cite{Heinz:2001xi,Hirano:2005xf,Huovinen:2001cy,Hirano:2002ds,Romatschke:2007mq,Luzum:2011mm,Song:2010mg,Qian:2016fpi,Magdy:2017ohf,Magdy:2017kji,Schenke:2011tv,Teaney:2012ke,Gardim:2012yp,Lacey:2013eia}.  
%

Experimentally azimuthal anisotropy of particle production relative to the reaction plane $\Psi_{R}$ can be characterized by the Fourier expansion~\cite{Poskanzer:1998yz} of the final-state azimuthal angle distribution,
\begin{eqnarray}
\label{eq:1-1}
\frac{dN}{d\phi} = \dfrac{N}{2\pi}  \left(  1+2\sum_{n=1}V_{n} e^{-in\phi} \right)  ,
\end{eqnarray}
where $V_{n} = v_{n}\exp(in\Psi_{n})$ is the n$^{\mathrm{th}}$ complex anisotropic flow vector and $\Psi_{n}$ and $v_{n}$ 
represent the vector direction and magnitude, respectively. First flow harmonic, $v_{1}$, is usually
referred to as  directed flow; $v_{2}$ is called elliptic flow, and $v_{3}$ the triangular flow, etc.
A large knowledge about  the properties of the matter created in heavy-ion collisions, has been gained through 
anisotropic flow studies of directed and elliptic flow~\cite{Magdy:2019ojv,Adam:2019woz,Magdy:2018itt}, higher-order flow harmonics $v_{n > 2}$ \cite{Adamczyk:2017ird,Magdy:2017kji,Adamczyk:2017hdl,Alver:2010gr, Chatrchyan:2013kba}, 
flow fluctuations \cite{Alver:2008zza,Alver:2010rt, Ollitrault:2009ie} and the correlation between 
different flow harmonics \cite{STAR:2018fpo,Adamczyk:2017hdl,Qiu:2011iv, Adare:2011tg, Aad:2014fla, Aad:2015lwa}.

Hydrodynamic models suggest that anisotropic flow arises from the evolution of the medium in the presence of initial-state anisotropies, defined by the eccentricities $\varepsilon_{n}$. The $v_{2}$ and $v_{3}$ flow harmonics observed to be,  linearly related to the initial-state anisotropies, $\varepsilon_{2}$ and $\varepsilon_{3}$, 
respectively \cite{Song:2010mg, Niemi:2012aj,Gardim:2014tya, Fu:2015wba,Holopainen:2010gz,Qin:2010pf,Qiu:2011iv,Gale:2012rq,Liu:2018hjh}. For these flow terms,
\begin{eqnarray}
\label{eq:1-2}
v_{n} = \kappa_{n} \varepsilon_{n},
\end{eqnarray}
where $\kappa_{n}$ is then encodes information about the medium properties such as the specific shear viscosity (the ratio of shear viscosity to entropy density) $\eta/s$.  Consequently, the experimental values of $v_{n} $ (for $n = 2, 3$) are routinely used as a constraint to probe $\eta/s$ for the QGP \cite{Adam:2019woz}. The higher-order $n \gt 3$ flow harmonics are of great interest as constraints for $\eta/s$ because they are more sensitive to the influence of viscous attenuation. However, they may not only arise from a linear response to the same-order initial-state anisotropies, but also from a nonlinear response to the lower-order 
eccentricities $\varepsilon_{2}$ and/or $\varepsilon_{3}$~\cite{Teaney:2012ke,Bhalerao:2014xra,Yan:2015jma}. Therefore, the full utility  of higher-order flow harmonics as constraints for $\eta/s$ extraction~\cite{Yan:2015jma} rests on robust 
separation of their linear (L) and nonlinear (nL) contributions.

The higher-order flow harmonics $V_{4}$ and $V_{5}$ can be expressed as:
\begin{eqnarray}\label{eq:1-3}
V_{4}  &=&  V_{4}^{\rm L} +  \chi_{4,22} V_{2} \, V_{2} \\
V_{5}  &=&  V_{5}^{\rm L} +  \chi_{5,23} V_{2} \, V_{3}, \nonumber
\end{eqnarray}
where  $\chi_{n, ij}$  represent the nonlinear response coefficients. The $V^{\rm nL}_{n > 3}$ encodes the 
correlations between different symmetry planes $\Psi_{n}$ which could helps to constrain the initial-stage dynamics \cite{Bilandzic:2013kga, Bhalerao:2014xra, Aad:2015lwa, ALICE:2016kpq, STAR:2018fpo,
Zhou:2016eiz, Qiu:2012uy,Teaney:2013dta, Niemi:2015qia, Zhou:2015eya}.

The magnitudes of $\chi_{4,22}$ and $\chi_{5,23}$ 
constrains the magnitude of $V^{\rm nL}_{n > 3}$ while $V^{\rm nL}_{n > 3}$ encodes the 
correlations between the flow symmetry planes $\Psi_{n}$ for different harmonic orders. The latter results from 
the initial-stage dynamics \cite{Bilandzic:2013kga, Bhalerao:2014xra, Aad:2015lwa, ALICE:2016kpq, STAR:2018fpo,
Zhou:2016eiz, Qiu:2012uy,Teaney:2013dta, Niemi:2015qia, Zhou:2015eya}.

The linear and nonlinear contributions to the  higher-order flow coefficients (Eq. \ref{eq:1-3}) can be easily separated if 
they are uncorrelated ({\em i.e.}, the linear mode is perpendicular to the nonlinear mode which is often assuumed)~\cite{Yan:2015jma,Qian:2016pau}. 
However, a test of the degree of a possible correlation between them constitutes an important prerequisite.
A common approach is to employ the Pearson's correlation coefficients for the higher and lower order flow harmonics~\cite{Bhalerao:2014xra,Bhalerao:2014xra,Acharya:2017zfg}:
\begin{eqnarray}
\frac{ \left< V_{4} \, (V_{2}^{*})^{2} \, v_{2}^{~2} \right>} {  \left< V_{4}  \, (V_{2}^{*})^{2} \right>  \, \left<v_{2}^{~2} \right> } \leftrightarrow  \frac{ \left< v_{2}^{~6} \right>} {  \left< v_{2}^{~4} \right>  \, \left<v_{2}^{~2} \right> } ,\label{eq:1-4} \\
\frac{ \left< V_{5} \, V_{3}^{*} \,  V_{2}^{*} \, v_{2}^{~2}  \right>} { \left< V_{5} \, V_{3}^{*} \,  V_{2}^{*}  \right> \, \left< v_{2}^{~2}\right>} \leftrightarrow  \frac{ \left< v_{2}^{~4} \, v_{3}^{~2} \right>} {  \left< v_{2}^{~2} \, v_{3}^{~2} \right>  \, \left<v_{2}^{~2} \right> }.
\label{eq:1-5}
\end{eqnarray}
Here, validation/invalidation of the independence between the linear and nonlinear modes of the higher-order flow harmonics
is obtained  by comparing both sides of Eqs. \ref{eq:1-4} and \ref{eq:1-5}.  
\begin{figure*}[ht]
\centering{
\includegraphics[width=1.0\linewidth,angle=0]{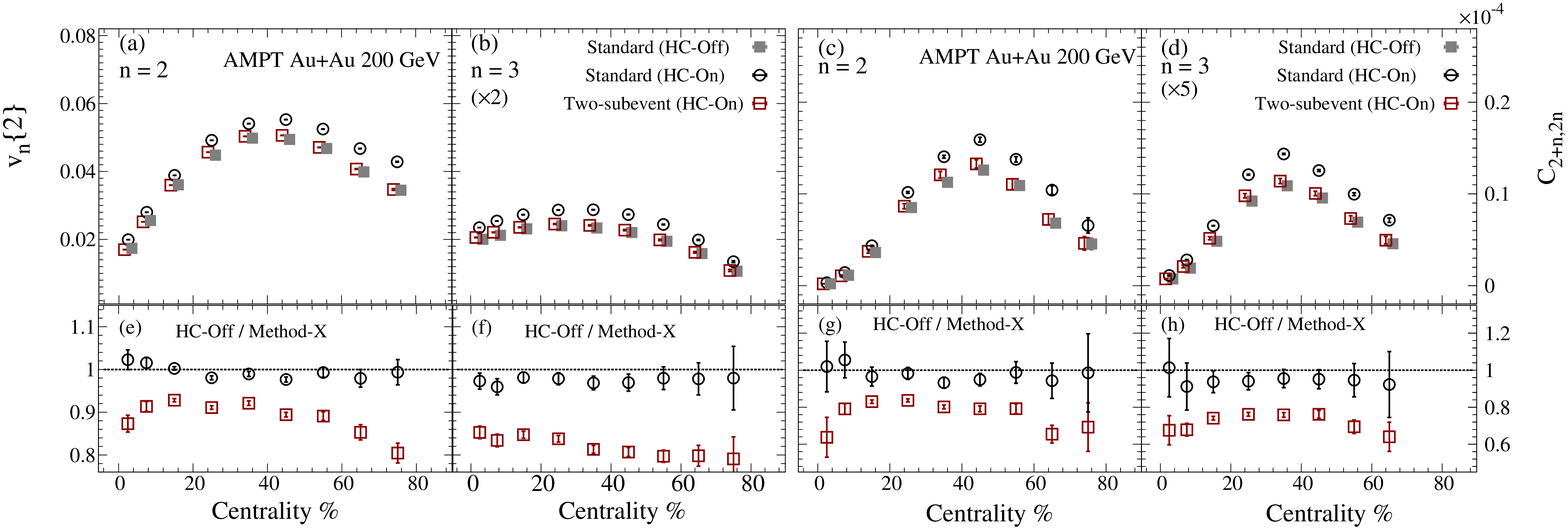}
\vskip -0.36cm
\caption{
The centrality dependence of the elliptic and triangular flow harmonics, $v_{2}$ and $v_{3}$, panels (a and b) and the three particle correlations, $C_{4,22}$ and $C_{5,23}$, (c and d) for Au+Au collisions at $\roots$ = 200 GeV from the AMPT model. The results are presented using the standard and the two subevent cumulant methods. 
The results are compared with similar simulations from the AMPT with the Hadronic cascade (H-C) mechanism turned off. The lower panels represent the ratios between the H-C off and the standard/two subevent cumulant methods.
\label{fig:fig1-2}
 }
}
\end{figure*}
An important caveat to the use of this method
is the demand for high statistical power ({\em i.e.}, large samples of events with sizable event multiplicity). 
Thus, it is of limited utility for carrying out checks for some experimental data, especially at lower beam energies.

In this work, we  investigate an alternative validation scheme which employs the Event-Shape Engineering technique (ESE)
to study both the linear and nonlinear modes of the higher-order flow harmonics.  Here, the underlying notion is that ESE not only gives access to more detailed differential measurements of the correlations between flow harmonics, but provides specific identifiable patterns for the strength of the coupling between the linear and nonlinear  modes of the higher-order flow terms. That is, for no-coupling,  the linear contributions to the higher-order flow harmonics should be event-shape independent while the nonlinear contributions should be event-shape dependent.

\section{Method}\label{Sec:2}

The current study is performed with simulated events for Au+Au collisions at $\sqrt{s_{NN}}$ = 200~GeV, obtained 
with the AMPT~\cite{Lin:2004en} model. For these events, the string melting mechanism in AMPT was turned on, 
and particles with transverse momentum $0.1 < p_T < 2.0$ \GeVc, were selected.
The model, which has been widely employed to study relativistic heavy-ion 
collisions \cite{Lin:2004en,Ma:2016fve,Ma:2013gga,Ma:2013uqa,Bzdak:2014dia,Nie:2018xog}, includes several important
ingredients: an initial parton conditions produced by the heavy ion jet interaction generator (HIJING) model~\cite{Wang:1991hta,Gyulassy:1994ew},
partonic interactions~\cite{Zhang:1997ej}, and conversion from partons to hadronic matter, followed by hadronic interactions~\cite{Li:1995pra}. 
The events produced by the AMPT model were analyzed with the multi-particle cumulant 
technique \cite{Bilandzic:2010jr,Bilandzic:2013kga,Jia:2017hbm,Gajdosova:2017fsc} in tandem with ESE.

The framework for the standard cumulant method is discussed in Refs.~\cite{Bilandzic:2010jr,Bilandzic:2013kga};
its extension to the subevents method is reported in Refs.~\cite{Jia:2017hbm,Gajdosova:2017fsc}. 
In the standard method, the $n^{\rm th}$-particle cumulants are constructed using particles from the fully available $\eta$ acceptance. 
Thus the constructed two-  and multi-particle correlations can be written as:
\begin{eqnarray}\label{eq:2-1}
v_{n} &=&  \langle  \langle \cos (n (\varphi_{1} -  \varphi_{2} )) \rangle  \rangle^{1/2},
\end{eqnarray}
\begin{eqnarray}\label{eq:2-2}
C_{k,n,m}           &=&   \langle \langle \cos ( k \varphi_{1} - n \varphi_{2} -  m \varphi_{3}) \rangle \rangle ,
\end{eqnarray}
\begin{eqnarray}\label{eq:2-3}
\langle v_{n}^{2} v_{m}^{2}  \rangle &=& \langle \langle \cos ( n \varphi_{1} + m \varphi_{2} -  n \varphi_{3} -  m \varphi_{4}) \rangle \rangle,
\end{eqnarray}
where, $\langle \langle \, \rangle \rangle$ represents the average over all particles in a single event, and then an average over all events, $k=n+m$, $n$ and $m$ are harmonic numbers and $\varphi_{i}$ expresses the azimuthal angle of the $i^{\rm th}$ particle.

To minimize the non-flow correlations resulting from resonance decays, Bose-Einstein correlation and fragments of individual jets to the $n^{\rm th}$-particle cumulants, that typically involve particles emitted within a localized region of rapidity, particles were grouped into two sub-events. Each sub-event covered a non-overlapping $\eta$-interval with  separation $|\Delta\eta| > 0.8$ between the sub-events $A$ and $B$ (i.e. $\eta_{A}~ > 0.4$ and $\eta_{B}~ < -0.4$) this $\Delta\eta$ separation takes into account the limited acceptance of the STAR detector~\citep{Magdy:2017ohf}. Here, the two-  and multi-particle correlations can be written as:
\begin{eqnarray}\label{eq:2-4}
v^{AB}_{n} &=&  \langle  \langle \cos (n (\varphi^{A}_{1} -  \varphi^{B}_{2} )) \rangle  \rangle^{1/2},
\end{eqnarray}
\begin{eqnarray}\label{eq:2-5}
C^{AB}_{k,n,m}           &=&   \langle \langle \cos ( k \varphi_{1}^{A} - n \varphi_{2}^{B} -  m \varphi_{3}^{B}) \rangle \rangle ,
\end{eqnarray}
\begin{eqnarray}\label{eq:2-6}
\langle v_{n}^{2} v_{m}^{2}  \rangle^{AB} &=& \langle \langle \cos ( n \varphi^{A}_{1} + m \varphi^{A}_{2} -  n \varphi^{B}_{3} -  m \varphi^{B}_{4}) \rangle \rangle.
\end{eqnarray}

These correlators were used to extract and study the linear and mode-coupled harmonics with event-shape selection.

\section{Results and discussion}\label{Sec:3}

The reliability of the extracted linear and mode-coupled harmonics can be influenced by possible short-range non-flow contributions to the two- and three-particle correlators used for the extractions. Therefore, it is instructive to evaluate a figure of merit for these contributions.
%
%
Figure~\ref{fig:fig1-2} compares the results obtained from the standard and two-subevents cumulant methods for simulated data with different conditions, specifically the Hadronic Cascade (HC) on and off options of AMPT. In AMPT the HC is based on the ART model ("A Relativistic Transport")~\citep{Li:1995pra,Li:2001xh} combining baryon-baryon, baryon-meson, and meson-meson elastic and inelastic scatterings. Using the HC-off option effectively eliminates a major part of non-flow contributions, and thus the effectiveness of the two sub-event approach for non-flow suppression could be tested.
The comparison of the standard and two sub-event cumulant methods results from HC-on sample shows larger $v_n$ magnitudes extracted via standard method than those from the two sub-events. Qualitatively, one expects to see such differences due to non-flow contributions. Further, the two sub-events method shows a good agreement (within 2\%) between HC-on and HC-off results, directly addressing this method ability to reduce correlation contributions from  non-flow effects.



Using  Eqs. \ref{eq:2-1} - \ref{eq:2-6}, the nonlinear mode in the higher order anisotropic flow harmonics 
of  $v_{4}$ and $v_{5}$ can be written as;
\begin{eqnarray}\label{eq:2-4}
v_{4}^{\rm nL} &=& \frac{C_{4,22}} {\sqrt{\langle \mathrm{v_2^2 v_2^2 }\rangle}}, \\ 
                                  &\sim &  \langle v_{4} \, \cos (4 \Psi_{4} - 2\Psi_{2} - 2\Psi_{2}) \rangle, \\ \nonumber
v_{5}^{\rm nL} &=& \frac{C_{5,23}} {\sqrt{\langle \mathrm{v_2^2 v_3^2 }\rangle}}, \\ 
                                  &\sim &  \langle v_{5} \, \cos (5 \Psi_{5} - 2\Psi_{2} - 3\Psi_{3}) \rangle, \\ \nonumber
\end{eqnarray}
and the linear contribution to $v_{4}$ and $v_{5}$ can be given as,
\begin{eqnarray}\label{eq:2-5}
v_{4}^{\rm L} = \sqrt{ (v_{4})^{\,2} - (v^{\rm nL}_{4})^{\,2}  }, \\ 
v_{5}^{\rm L} = \sqrt{ (v_{5})^{\,2} - (v^{\rm nL}_{5})^{\,2}  }.  \nonumber
\end{eqnarray}
Equation (\ref{eq:2-5}) assumes that the linear and nonlinear contributions in $v_{4}$ and $v_{5}$ are independent~\cite{Yan:2015jma}, which is only valid if the correlations between the $v_{n}$ ($n=2,3$) and the higher-order flow coefficients ($n>3$) is weak. 

The degree of this correlation can be tested via ESE via selections on  the magnitude of the second-order reduced flow vector $q_{2}$ \cite{Adler:2002pu}:
\begin{eqnarray}
Q_{2, x} = \sum_{i} \cos(2 \varphi_{i}),
Q_{2, y} = \sum_{i} \sin(2 \varphi_{i}),
\end{eqnarray}
\begin{eqnarray}
q_{2}    &=& \frac{|{Q}_{2}|}{\sqrt{M}}, ~|Q_{2}|  = \sqrt{Q_{2, x}^2 + Q_{2, y}^2}
\end{eqnarray}
where $Q_{2}$ is the magnitude of the second-order harmonic flow vector calculated from the azimuthal 
distribution of particles within $|\eta| < 0.3$, and $M$ is the charged hadron multiplicity of the same sub-event. Note that the associated flow measurements are performed within $|\eta| > 0.4$ which allows for a separation between the $q_{2}$ subevent and the flow measurements subevents.

%
\begin{figure}[t]
\centering{
\includegraphics[width=1.02 \linewidth, angle=0]{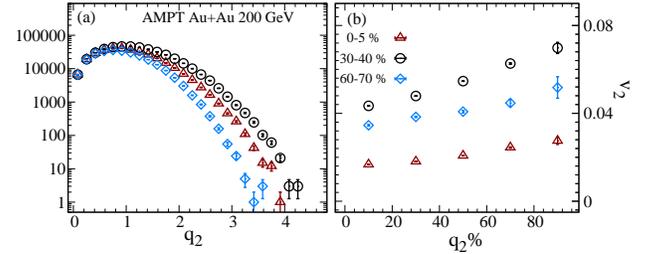}
\vskip -0.46cm
\caption{
 The $q_{2}$ distribution for Au+Au collisions at $\sqrt{s_{NN}}$ = 200~GeV in the centrality classes $0-5\%$, $30-40\%$ and $60-70\%$, for the sub-event sample with  $\mathrm{|\eta| < 0.3}$ shown in panel (a). An illustrative plot of $ v_{2}$ as a function of the $q_{2}$ percentile selections are shown in (b).
\label{fig:fig2}
 }
}
\end{figure}
%
\begin{figure}[t]
\centering{
\includegraphics[width=1.02 \linewidth, angle=0]{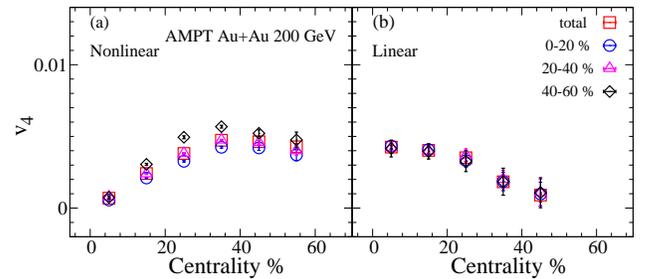}
\vskip -0.46cm
\caption{
The linear and nonlinear $v_{4}$ using the two subevent cumulant method as a function of centrality for different event-shape selections from the AMPT model are shown.
\label{fig:fig3}
 }
}
\end{figure}
%
\begin{figure}[t]
\centering{
\includegraphics[width=1.02 \linewidth, angle=0]{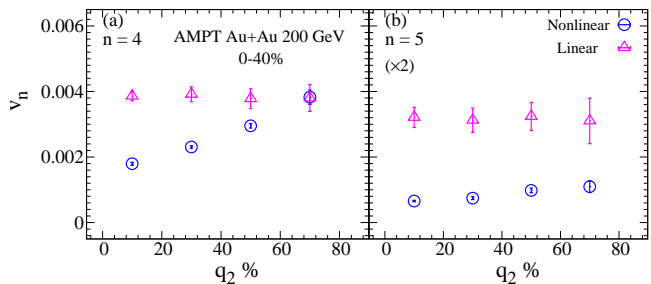}
\vskip -0.46cm
\caption{
The linear and nonlinear $v_{4}$ and $v_{5}$ using the two subevent cumulant method as a function of $q_{2}\%$ for centrality selection $0-40\%$ from the AMPT model are shown.
\label{fig:fig4}
 }
}
\end{figure}
%

Figure~\ref{fig:fig2} (a) shows that the patterns of the $q_{2}$ distributions for several centrality selections, are similar. Fig.~\ref{fig:fig2}~(b) demonstrates the linear correlation of these $q_{2}$ selections on the magnitude of $v_{2}$ for several centrality selections. Here, it is noteworthy that the sensitivity of the ESE method, and hence, its effectiveness as a tool, depends on the multiplicity and the magnitude of $v_{2}$, both of which decreases with beam energy~\citep{Bzdak:2019pkr}. 
Moreover, the non-flow effects (i.e. resonance decays, jets, etc.~\citep{Voloshin:2008dg}) could bias the $q_{2}$ selections. 
The bias, however, could be mitigated by using the subevents cumulant methods~\cite{Jia:2017hbm,Gajdosova:2017fsc}.

The centrality dependence of the linear and nonlinear $v_{4}$ is shown in Fig.~\ref{fig:fig3} for 0-20\%, 20-40\% and 40-60\% $q_{2}$ selections . The results indicate  that the nonlinear mode of $v_{4}$ depends strongly on both collision centrality and $q_{2}$ selection. This dependence reflects the dominant role of $v_{2}$ contributions to this nonlinear mode.
By contrast, the linear mode of $v_{4}$ also shows a strong dependence on centrality but little, if any, dependence on $q_2$. 
This observation can be attributed to the weak coupling between the linear and nonlinear modes of $v_{4}$ in the AMPT.

The  linear and nonlinear modes for $v_{4}$ and $v_{5}$ for 0-40\% central events are presented 
as a function of $q_{2}$ in Fig.~\ref{fig:fig4}. They further indicate that while $v^{\rm nL}_{4}$ and $v^{\rm nL}_{5}$ 
increase with the  $q_{2}$ selection, the liner mode of $v_{4}$ and $v_{5}$ shows no $q_{2}$ dependence,
further suggesting a weak correlation between the linear and nonlinear modes of $v_{4}$ and $v_{5}$.

\section{Summary}\label{Sec:4}

In summary, we studied the event shape sensitivity of the linear and mode-coupled contributions to the $v_4$ and $v_5$ anisotropic flows in the AMPT model using Event-Shape Engineering. The nonlinear contribution to these higher-order flow harmonics is found to be event-shape dependent while the linear contribution showed little, if any, event-shape sensitivity, reflecting no measurable correlation between the linear and nonlinear contributions to $v_4$ and $v_5$ in AMPT. These observations illustrate that event-shape selections can be used to examine the degree of correlation between the linear and nonlinear contributions to the higher-order flow harmonics and consequently, aid reliable extractions of the linear and mode-coupled $v_n$ required to constrain precision extractions of $\eta/s(\mu_B,T)$.

\section*{Acknowledgments}
%
The authors thank Emily E. Racow for the useful discussion.
This research is supported by the US Department of Energy under contract DE-FG02-94ER40865 (NM and OE) and DE-FG02-87ER40331.A008 (RL).
%
%
\bibliography{ref} 
\end{document}